\documentstyle[12pt,twoside,fleqn,espcrc1,psfig]{article}

\newcommand{\AmS}{{\protect\the\textfont2
  A\kern-.1667em\lower.5ex\hbox{M}\kern-.125emS}}
\title{Uncertainties in direct neutron capture calculations
due to nuclear structure models}

\author{T. Rauscher\address{Institut f\"ur Physik, 
        Universit\"at Basel, \\ 
        Klingelbergstr.\ 82, CH-4056 Basel, Switzerland}%
        \thanks{APART fellow of the Austrian Academy of Sciences},
K.-L. Kratz\address{Institut f\"ur Kernchemie, Universit\"at Mainz,\\
Fritz-Strassmann-Weg 2, D-55099 Mainz, Germany},
H. Oberhummer\address{Institut f\"ur Kernphysik, TU Wien,\\
Wiedner Hauptstr.\ 8--10, A-1040 Wien (Vienna), Austria},
J. Dobaczewski\address{Institute for Theoretical Physics,
Warsaw University,\\
ul.\ Hoza 69, 00-681 Warsaw, Poland}, 
P. M\"oller\address{P. Moller Scientific Computing and Graphics Inc.,\\
Los Alamos, New Mexico, USA}, and
M. Sharma\address{Dept.\ of Physics, Kuwait University\\
Kuwait 13060}}

\begin{document}
\maketitle

\begin{abstract}
The prediction of cross sections for nuclei far off stability is crucial
in the field of nuclear astrophysics. For spherical nuclei close to the
dripline the statistical model (Hauser-Feshbach) approach is not
applicable and direct contributions may dominate the
cross sections. For neutron-rich, even-even Sn targets, we compare the 
resulting 
neutron capture cross sections when consistently
taking the input for the direct capture calculations from three different
microscopic models. The results
underline the sensitivity of cross sections calculated in the direct model
to nuclear structure models which can lead to high uncertainties when
lacking experimental information.
\end{abstract}

\section{INTRODUCTION}

\begin{figure}
\psfig{file=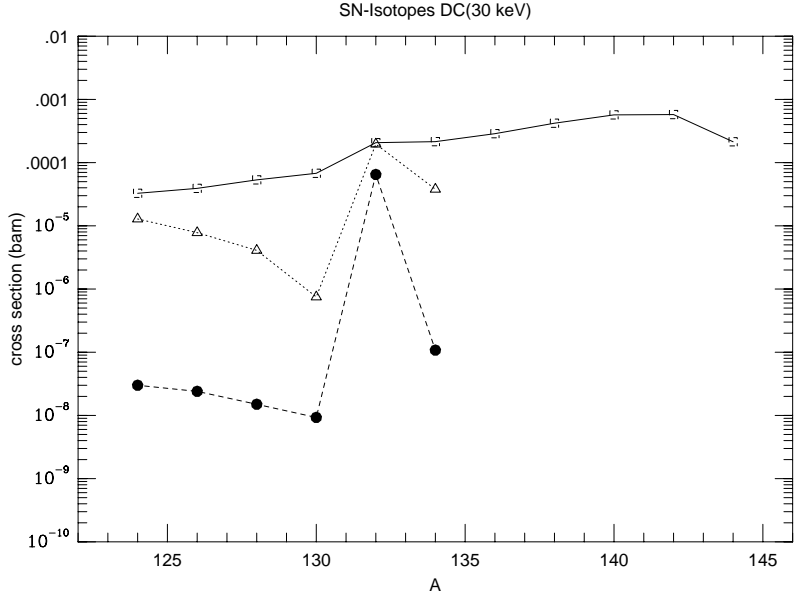,height=9.53cm,rheight=8.3cm}
\caption{\label{snfig}\small Direct-capture cross sections at 30
keV for different Sn isotopes.
Levels and masses are calculated with models by Sharma et
al.~\protect{\cite{sharma}} (triangles),
M\"oller et al.~\protect{\cite{moeller}} (dots), and Dobaczewski et
al.~\protect{\cite{doba}} (squares). The lines are drawn to guide the eye.}
\end{figure}

\begin{figure}
\psfig{file=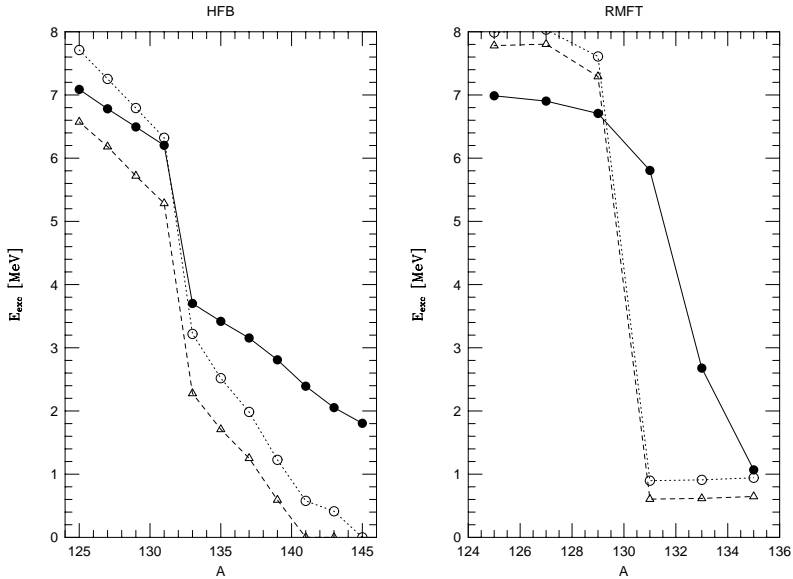,height=9.53cm,rheight=8.3cm}
\caption{\label{states}\small Dependence of level energies on mass number for
the even-odd isotopes $^{125-135}$Sn in the RMFT model~\protect{\cite{sharma}}
(right) and for the isotopes $^{125-145}$Sn in the HFB 
model~\protect{\cite{doba}} (left).
Shown are the 1/2$^-$ state (open circles), the
3/2$^-$ state (triangles) and the calculated neutron separation energy (full
circles). The lines are drawn to guide the eye. Note the different range in
mass numbers in the two plots.}
\end{figure}

Explosive nuclear burning in astrophysical environments produces
unstable nuclei which again can be targets for subsequent reactions.
Most of these nuclei are not accessible in terrestrial labs or not fully
explored by experiments, yet. For the majority of unstable nuclei the
statistical model (Hauser-Feshbach) can be used to determine the cross
sections. However, for nuclei close to the dripline the level density
becomes too low to apply the statistical model~\cite{tommy}
and contributions of a direct
interaction mechanism (DI) may dominate the cross sections. 
The DI requires the detailed knowledge of energy levels (excitation energies,
spins, parities), contrary to the statistical model which averages over
resonances. Lacking experimental data, this information has to be extracted
from microscopic nuclear-structure models.

We compare the results for direct neutron capture (calculated in the optical
model~\cite{kim}) on the even-even isotopes
$^{124-145}$Sn with energy
levels,
masses, and nuclear density distributions
taken from different nuclear-structure models.
The utilized structure models were a Hartree-Fock-Bogoliubov model
(HFB) with SkP force~\cite{doba}, a Relativistic Mean Field Theory (RMFT) 
with the 
parameter set NLSH~\cite{sharma} and a Shell Model based on folded-Yukawa
wave functions (FYSM)~\cite{moeller}. 

A similar study has already been performed for neutron-rich Pb
isotopes~\cite{tom}.

\section{METHOD}

The cross sections were calculated in the optical model for direct
capture~\cite{kim}, utilizing optical potentials derived by the folding
procedure~\cite{satch}. In the folding approach the nuclear target density
is folded with an energy- and density-dependent effective nucleon-nucleon
interaction in order to obtain the potentials for the bound and
scattering states. Only one open parameter $\lambda$ remains which accounts
for the effects of antisymmetrization and is close to unity.
The densities required for the determination of the
folding potentials were consistently calculated from the wave functions
of the respective nuclear-structure model.
For the bound states the strength parameter $\lambda$ was fixed by the condition
to reproduce the given binding energy of the captured neutron. The 
value of $\lambda$ for the scattering potential was adjusted to yield the
same value of 425 MeV fm$^3$~\cite{werner} for the volume integral as 
determined from the
experimental scattering data on stable Sn isotopes~\cite{mug,cinda}.

In order to be able to directly
compare the different models, all nuclei were assumed to be spherical and
the spectroscopic factors were set to 1.

\section{RESULTS AND DISCUSSION}

The results of the calculations for projectiles at $E_{\mathrm{c.m.}}=30$ keV
are summarized in Fig.~\ref{snfig}. For each model we calculated the 
capture cross section only up to the r-process path. The most extreme
location of the
path (farthest away from the line of stability) is determined by neutron
separation energies $E_{\mathrm{n}} \approx 2$ MeV~\cite{fkt}. Depending on
the microscopic model, the path will then be located at higher or lower
mass numbers $A$. In the case of RMFT and FYSM it will go through $A\approx$
132--134, for HFB the path will be shifted to
considerably higher mass numbers $A\approx$ 142--144. (The neutron dripline
is also shifted to higher masses in the latter model.)

Similar effects as seen in the behavior of the Pb cross sections~\cite{tom} 
can also be found for the Sn cross sections. The cross section can vary by
order of magnitudes when going from one isotope to the next and also differ
vastly between the different microscopic models.

As the capture to low-spin states ($J$=1/2, 3/2) accounts for the largest
contributions to the cross section, the results are very sensitive to the
presence of bound low-spin states. Since the microscopic models not only
yield different masses (i.e.\ neutron separation energies) but also exhibit
different behaviors of the level energies with changing mass,
``jumps'' and ``gaps'' can be seen with some models (RMFT, FYSM), whereas 
others (HFB)
result in a smoother behavior of the capture cross sections in an 
isotopic chain.
This is illustrated in Fig.~\ref{states}, which shows the neutron
separation energy and the excitation energy of the 1/2$^-$ and 3/2$^-$
states in RMFT and HFB. As long as both states are unbound in the RMFT, the
cross sections remain low and only jump to higher values when those
states become bound at the shell closure. As at least the 3/2$^-$ level is
always bound in HFB, the cross sections show a smoother behavior.
The variation in the FYSM cross sections can be explained in a similar
way.

\section{CONCLUSION}

With this work we have underlined that the calculation of purely
theoretical direct capture cross sections far from stability still contains a
large error, even when using most recent nuclear-structure models.
In the previously discussed case of Pb isotopes~\cite{tom}, the r-process path 
contains nuclei in or at the border of a region expected to be deformed, 
leading to
higher level densities and thus favoring the compound nucleus mechanism.
This is not true for neutron-rich isotopes in the Sn region, especially
around the neutron magic number $N=82$ where the level density becomes too
low for the statistical model. Therefore the neutron capture cross sections
have to be calculated using input from nuclear-structure models and will
be subject to the quoted uncertainties, even when the different models
yield similar values for other nuclear properties, such as masses.

Similar problems may be encountered on the proton-rich side when predicting
proton capture cross sections close to the proton dripline.

\end{document}